\documentclass{article}
\usepackage[preprint]{spconfa4}
\usepackage{amsmath,graphicx,amssymb,subfigure,svg}

\copyrightnotice{978-1-6654-6867-1/22/\$31.00~\copyright2022 IEEE}

\title{ACOUSTIC Room COMPENSATION USING Local PCA-BASED ROOM AVERAGE Power Response ESTIMATION }
\name{Wenyu Jin, Patrick McPherson, Chris Pike, Adib Mehrabi}
\address{Sonos, Inc., Boston, MA 02111, United States}
\begin{document}
\ninept
\def\baselinestretch{.903}\let\normalsize\small\normalsize
\maketitle
\begin{abstract}
Acoustic room compensation techniques, which allow a sound reproduction system to counteract undesired alteration to the sound scene due to excessive room resonances, have been extensively studied. Extensive efforts have been reported to enlarge the region over which room equalization is effective and to contrast variations of room transfer functions in space. A speaker-tuning technology ``Trueplay" allows users to compensate for undesired room effects over an extended listening area based on a spatially averaged power response of the room, which is conventionally measured using microphones on portable devices when users move around the room. In this work, we propose a novel system that leverages measured speaker echo path self-responses to predict the room average power responses using a local PCA based approach. Experimental results confirm the effectiveness of  the proposed estimation method, which further leads to a room compensation filter design that achieves a good sound similarity compared to the reference system with the ground-truth room average power response while outperforming other systems that do not use the proposed estimator.
\end{abstract}
\section{Introduction}
\label{sec:intro}
When sound is reproduced by a sounding unit, e.g. a loudspeaker, the perception of the sound that is delivered to the listener is modified by the listening environment. Excessive reflections or resonances may result in detrimental impacts on the original sound, e.g. altered music timbre, incorrect sound localization. The sound reproduction system can also introduce undesired artifacts, e.g. frequency band-extension, non-linearities and etc..

Room response compensation (RRC) has been widely studied \cite{karjalainen2005room,karjalainen2001low,karjalainen2002frequency,mourjopoulos2003comments,Jin13,Jin15} for improving sound reproduction quality, contrasting detrimental effects of the room environment. In an RRC system, the room transfer function (RTF) characterizing the path from the sound reproduction system to the listener is equalized with a suitably designed equalizer through its inversion. From \cite{app8010016}, RRC approaches are categorized into single-point and multi-point room equalizers. A single-point room equalizer derives the equalization filter based on the measurement in a single location of the RTF \cite{neely1979invertibility}, which is effective only in a limited zone around the measured point. In real-world scenarios, RTFs vary significantly with respect to the position in the room \cite{mourjopoulos1985variation} and time \cite{hatziantoniou2004errors,Jin16}. To enlarge the equalization zone and to contrast the room response variations, multi-point equalizers have been proposed \cite{bharitkar2006immersive}. A multi-point room equalization system leverages multiple measurements of RTFs at different locations in order to design the equalizer in a either fixed or adaptive manner. However, the placement and measurements at numerous separate locations that covers the extended room area could be a time-consuming process.

Alternatively, RTFs can be estimated from a finite set of room impulse response measurements at specific positions in the room through model-based approaches. Several works have been reported in the literature, such as common-acoustical pole and residue modelling \cite{Haneda99}, compressive sensing \cite{Mignot13,Verburg18} and the planewave decomposition method \cite{Jin14,Antonello17}. However, depending on the formulation of sparsity models that are used to characterize the entire soundfield in a room, these approaches still require a relatively large number of microphone measurements to achieve sufficient accuracy and microphone positions in space are also required. The estimation of RTFs is often severely under-determined, which is typically due to the large number of independent parameters needed to represent the acoustic system in reverberant environments. To this end, Fozunbal et al. \cite{Fozunbal08multi-channelecho} deployed Principal Component Analysis (PCA) and proposed a supervised algorithm by forming a model based on a training set that consists of pre-measured acoustic impulse responses from several known locations. Koren et al. \cite{Koren12} further improved the PCA-based estimation approach by exploiting spatial information of localized sources in order to properly train local models that corresponds to different regions in the room. PCA is also applied in \cite{Jin22} to facilitate the estimation of the sound pressure at the eardrum using an inward-facing microphone at the in-ear earpiece.

An efficient variant of the multi-point equalization system can be formulated based on a spatially-averaged power response measurement. While a power measurement necessarily eliminates phase information, such "room curves" are commonly used by speaker designers and reveal problems that are common to all locations, such as boundary effects \cite{TooleSoundReproduction}. Sheen \cite{SheenTrueplay} proposed a speaker tuning technology ``Trueplay" \cite{Trueplay} based on room power response measurements. The room average power response is typically derived using microphones on portable devices such as mobile phones. The user moves around the room with the portable device while the speaker under test plays a cyclic stimulus signal. The power response of each cycle is averaged to create the room average power response.

In this paper we present a novel supervised approach to estimate the room average power response by forming models based on a training set. The system requires only an on-board microphone while eliminating the need for additional microphone measurements and the knowledge of speaker position in space. The proposed room compensation design uses the echo path, i.e. the acoustic impulse response between the speaker and on-board microphone, to predict the room average power response. Specifically, the echo path measurement is projected onto a lower dimensional space using PCA and transformed to obtain an estimate of the room average power response. It also exploits local features of echo path self-responses (e.g. reverberation time) and divides the global training dataset into local subsets, thereby deriving separate local PCA models that corresponds to different room conditions. Results demonstrate the improved performance of the local PCA approach over other conventional estimation schemes and when this estimate is incorporated into the design of ``Trueplay" for room compensation, the proposed approach achieves good similarity to the reference system with the ground-truth room average power response.
\section{Problem Formulation}
\label{sec:2}
\begin{figure}
\centering
\includegraphics[width=0.65\linewidth]{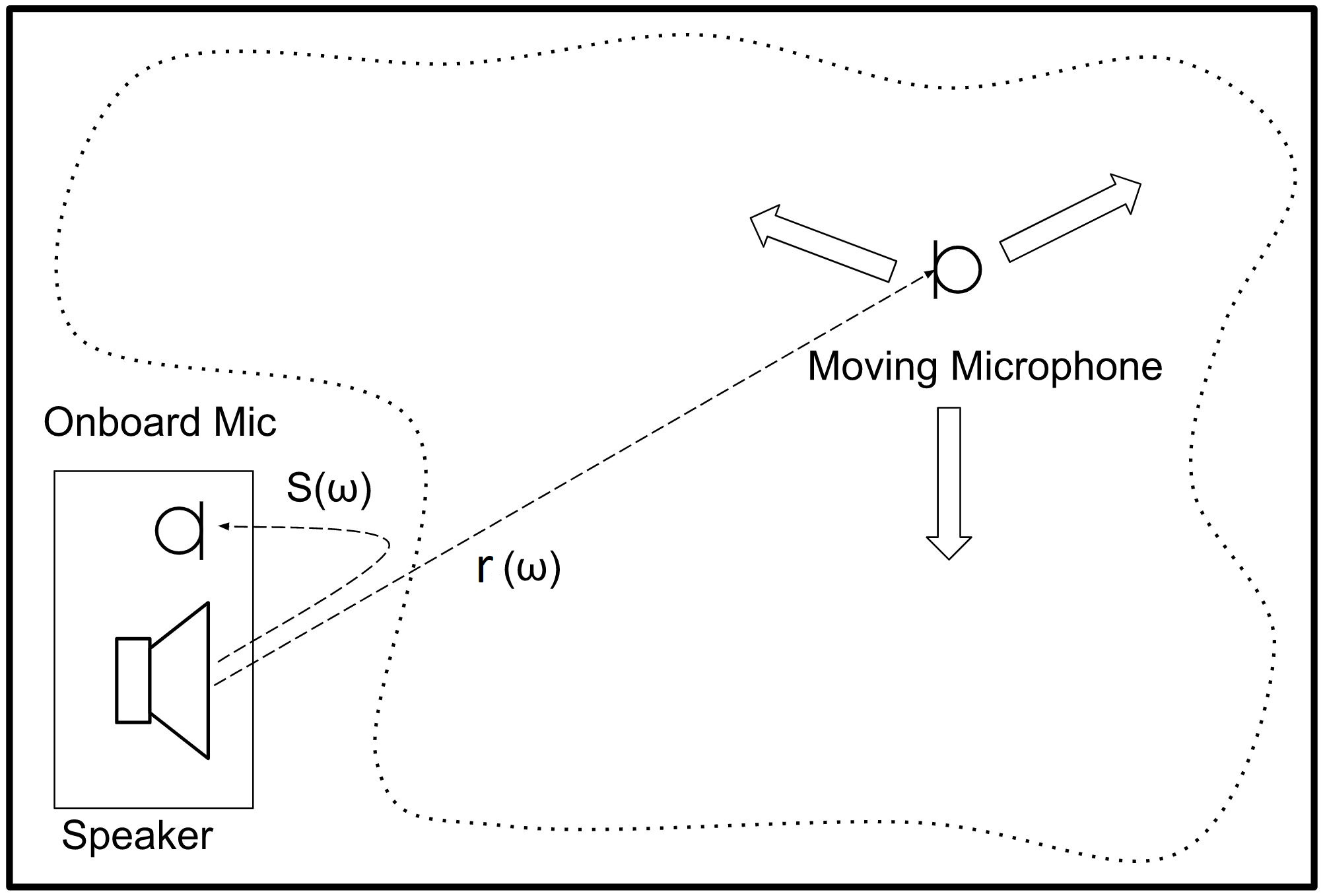}
\vspace{-0.2cm}
\caption{\label{fig:FIG1}Loudspeaker room compensation acoustic setup.}
\vspace{-0.3cm}
\end{figure}
The acoustic setup of ``Trueplay" is depicted in
Fig.~\ref{fig:FIG1}. The room average power response $r(\omega)$ can be measured at radial frequency $\omega$ using an external microphone as described in Sec.~\ref{sec:intro}. The on-board microphone measures the echo path self-response $s(\omega)$ of the device. The goal is to derive an equalization filter $G_{eq}(\omega)$ such that the room average power response meets a target magnitude response $t(\omega)$:
\begin{equation}
\vspace{-0.1cm}
r(\omega)|G_{eq}(\omega)|^2 = t(\omega).
\end{equation}
The target response is a subjective choice made by the loudspeaker designer. Targets are commonly flat across frequency with some boosts in the bass portion of the response. The magnitude response of the ideal filter is then given by
\begin{equation}
\vspace{-0.1cm}
|G_{eq}(\omega)| = \sqrt{\frac{t(\omega)}{r(\omega)}}.
\label{eq:EQ0}
\vspace{-0.1cm}
\end{equation}
The magnitude response is then turned into a minimum phase finite impulse response (FIR) filter by taking it's real cepstrum \cite{PeiCepstrum2006}, providing a realizable filter. The FIR can be used directly for equalization. To reduce computational complexity, the Trueplay system instead computes a set of second order infinite impulse response (IIR) filters that approximate the magnitude and phase response of the "ideal" FIR. Various techniques exist for the design of arbitrary IIR filters \cite{dodds2020a}\cite{lee2008simple}.

The main challenge of this paper is to replace $r(\omega)$ in Eq.~\eqref{eq:EQ0} with an appropriate estimator of the room average power response $\hat{r}(\omega)$ based on the in-situ measurement of the echo path self-response, which will be described in the following section.
\section{Estimation of room average power response}
\label{sec:3}
In this section, we present an estimation scheme to map the measured echo path self-response at the on-board microphone to the room average power response. The first subsection introduces a linear least-sqaures (LS) regression method that minimizes mean-squared error of the estimated room average power response coefficients in the frequency domain. Subsequently, we propose an estimation method that benefits from the numerical robustness and efficiency of PCA. Finally, we motivate a local PCA based method that trains individual local models corresponding to different room reverberant conditions.
\subsection{\label{subsec:3:1} Linear least-squares regression}
Let $\mathcal{M}=\{\mathbf{s_{j}}, \mathbf{r_{j}} \in \mathbb{R}^{\frac{N_f}{2}+1} | j=1,\ldots,J\}$ be a set of magnitude spectrum coefficients of echo path self-responses $s_{j}(\omega)$ and room average power responses $r_{j}(\omega)$ in the log-energy form. $N_f$ is the FFT length. The optimal filter $\mathbf{g_{LS}}$ should minimize the difference between the estimated room average power response to the measured counterpart via linear mapping in the frequency domain, i.e., the following least-squares cost function
\begin{equation}
E(\mathbf{g_{LS}})=\|\mathbf{D_{s}}\mathbf{g_{LS}}-\mathbf{d_{r}}\|^{2}_{2}+\mu\|\mathbf{g_{LS}}\|^{2}_{2},\label{eq:EQ1}
\end{equation}
where $\mathbf{D_{s}}$ ($J(\frac{N_f}{2}+1) \times (\frac{N_f}{2}+1)$) are stacked diagonal matrices containing magnitude coefficients of all measured echo path self-responses $s_{j}(\omega)$ and $\mathbf{d_{r}}$ ($J(\frac{N_f}{2}+1) \times 1$) is the stacked vectors containing all $r_{j}(\omega)$. The Tikhonov regularization factor $\mu=0.001$ is considered in this work to prevent over-amplification of $\mathbf{g_{LS}}$. The optimum with respect to $\mathbf{g_{LS}}$ is given by
\begin{equation}
\mathbf{\hat{g}_{LS}}=(\mathbf{D_{s}}^{H}\mathbf{D_{s}}+\mu \mathbf{I})^{-1}\mathbf{D_{s}}^{H}\mathbf{d_{r}}.
\end{equation}
Using the estimated filter $\mathbf{\hat{g}_{LS}}$ from the training stage, room average power response coefficients $\mathbf{\hat{r}_{LS}}$ are estimated from a measurement of the echo path self-response $\mathbf{s}$ during run-time as follows: 
\begin{equation}
\mathbf{\hat{r}_{LS}}=\mathbf{s} \odot \mathbf{\hat{g}_{LS}}, \label{eq:EQ5}
\end{equation}
where $\odot$ denotes element-wise product.
\subsection{\label{subsec:3:2} Global PCA-based estimation}
As can be seen from \eqref{eq:EQ1}, a direct linear mapping of the frequency domain vectors $\mathbf{s_{j}}, \mathbf{r_{j}}$ is possible but would require a large set of training data to ensure sufficient accuracy if the FFT length is large. In this section, a PCA-based estimator of $\mathbf{r}$ is designed based on features of measured echo path self-responses $\mathbf{s}$.

PCA is commonly used for dimensionality reduction by projecting each data point onto only the first few principal components to obtain lower-dimensional data that helps to avoid over-fitting \cite{Steward93}. By conducting PCA, we extract the first $K_s$ and $K_r$ principal components $\mathbf{u}_{s,k},\mathbf{u}_{r,k} \in \mathbb{R}^{\frac{N_f}{2}+1}$ of $\mathbf{s_{j}}$ and $\mathbf{r_{j}}$, respectively. The room average power response principal components matrix is defined as
\begin{equation}
\mathbf{U}_{r}= [\mathbf{u}_{r,1},\mathbf{u}_{r,2},\ldots,\mathbf{u}_{r,K_{r}}].
\end{equation}
Let $\mathbf{\bar{r}}$ be the ensemble average of $\mathbf{r_{j}}$: $\mathbf{\bar{r}}=\sum_{j=1}^{J} \mathbf{r_{j}}/J$. To obtain gain vectors $\mathbf{g}_{r,j}$ that minimize the Euclidean distance between the reconstructed frequency domain vectors
\begin{equation}
\mathbf{\hat{r}}_{j}= \mathbf{\bar{r}}+\mathbf{U}_{r}\mathbf{g}_{r,j}
\end{equation}
and the raw $\mathbf{r}_{j}$, we utilize the orthonormality of the principal components and get
\begin{equation}
\mathbf{g}_{r,j}=\mathbf{U}_{r}^{H}(\mathbf{r}_{j}-\mathbf{\bar{r}}).
\end{equation}
Similarly we obtain the gains $\mathbf{g}_{s,j}$ for the echo path.

After converting frequency domain coefficients into the principal component domain, the problem is to find a linear map $\mathbf{A} \in \mathbb{R}^{K_{s} \times K_{r}}$ that projects the echo path self-response gain vectors onto the room average power response gain vectors. The following cost function is defined:
\begin{equation}
\vspace{-0.2cm}
C(\mathbf{A})= \sum_{j=1}^{J}  \|\mathbf{g}_{r,j}-\mathbf{A}\mathbf{g}_{s,j}\|^{2}.\label{eq:EQ2}
\vspace{-0.1cm}
\end{equation}
This projection
allows us to estimate the room average power response gain vector given the echo path self-response gain vector. To minimize $C(\mathbf{A})$, we have
\begin{equation}
\vspace{-0.2cm}
\mathbf{\hat{A}}= \underset{\mathbf{A}}{\mbox{argmin}} \ C(\mathbf{A})=\sum_{j=1}^{J} \mathbf{g}_{r,j}\mathbf{g}_{s,j}^{H} (\sum_{j=1}^{J} \mathbf{g}_{s,j}\mathbf{g}_{s,j}^{H})^{-1}.
\vspace{-0.1cm}
\label{eq:EQ3}
\end{equation}
The above-demonstrated steps are the training stage for room average power response estimation based on a training set $\mathbf{s_{j}}, \mathbf{r_{j}} \in \mathcal{M}$. After measuring the echo path self-response $\mathbf{s}$ at runtime, the gain
vector for $\mathbf{s}$ in the the principal component domain can be calculated as follows:
\begin{equation}
\mathbf{g}_{s}=\mathbf{U}_{s}^{H}(\mathbf{s}-\mathbf{\bar{s}}),
\end{equation}
where $\mathbf{\bar{s}}$ is the ensemble average of $\mathbf{s}_{j}$ from the training stage. Then an estimate of $\mathbf{g}_{r}$ can be obtained by $\mathbf{\hat{g}}_{r}=\mathbf{\hat{A}}\mathbf{g}_{s}$. Finally an estimate $\mathbf{\hat{r}_{pca}}$ for $\mathbf{r}$ in the frequency domain with the ensemble average $\mathbf{\bar{r}}$ and $\mathbf{U}_{r}$ from the training stage
\begin{equation}
\mathbf{\hat{r}_{pca}}=\mathbf{\bar{r}}+\mathbf{U}_{r}\mathbf{\hat{g}}_{r}. \label{eq:EQ6} 
\end{equation}
\subsection{\label{subsec:3:3} Local PCA-based estimation}
 The PCA-based estimator of the room average power response in Sec.~\ref{subsec:3:2} is derived based on the global training dataset that consists of a mixture of measured echo path self-response/room average power response sets under a large variety of room conditions. In this section, we propose to exploit local features of echo path self-responses to ``cluster" the global training dataset into small subsets and to train local PCA models that corresponds to different room conditions accordingly, which facilitates PCA to better capture the structure of acoustic responses and their variation patterns.
 
 The following features that divide the global dataset into sub-groups are investigated in the training stage:
 \begin{itemize}
  \item Reverberation time RT30 \cite{Schroeder65} of the measured echo path self-response to quantify the reverberant condition at the position where the speaker is placed;
  \item Low-frequency roll-off gain of the echo path frequency response. In this work, we choose to monitor the magnitude roll-off between 120 Hz and 60 Hz.
\end{itemize}
At the inference stage, the selection criteria that indicates which local PCA model to be used for deriving an estimated room average power response $\mathbf{\hat{r}_{lpca}}$ when a new echo path self-response becomes available is also established accordingly. The effectiveness of the proposed local PCA estimation approach is evaluated in Sec. 4.
\vspace{-0.2cm}
\section{\label{sec:4}System Evaluation}
In this section, we evaluate the effectiveness of incorporating the proposed room average power response estimator into the "Trueplay" room compensation filter design. First the acoustic setup and the database of measurements are introduced. Second, we present results that demonstrate the estimation accuracy of the proposed PCA-based approaches. Finally, results of a perceptual evaluation of the proposed room compensation design  utilizing the multi stimulus hidden reference and anchor (MUSHRA) framework are presented.
\vspace{-0.2cm}
\subsection{\label{subsec:4:1}Self-response/Room average power response database}
\begin{figure}[t]
\centering  
\subfigure{
\includegraphics[width=.5\linewidth]{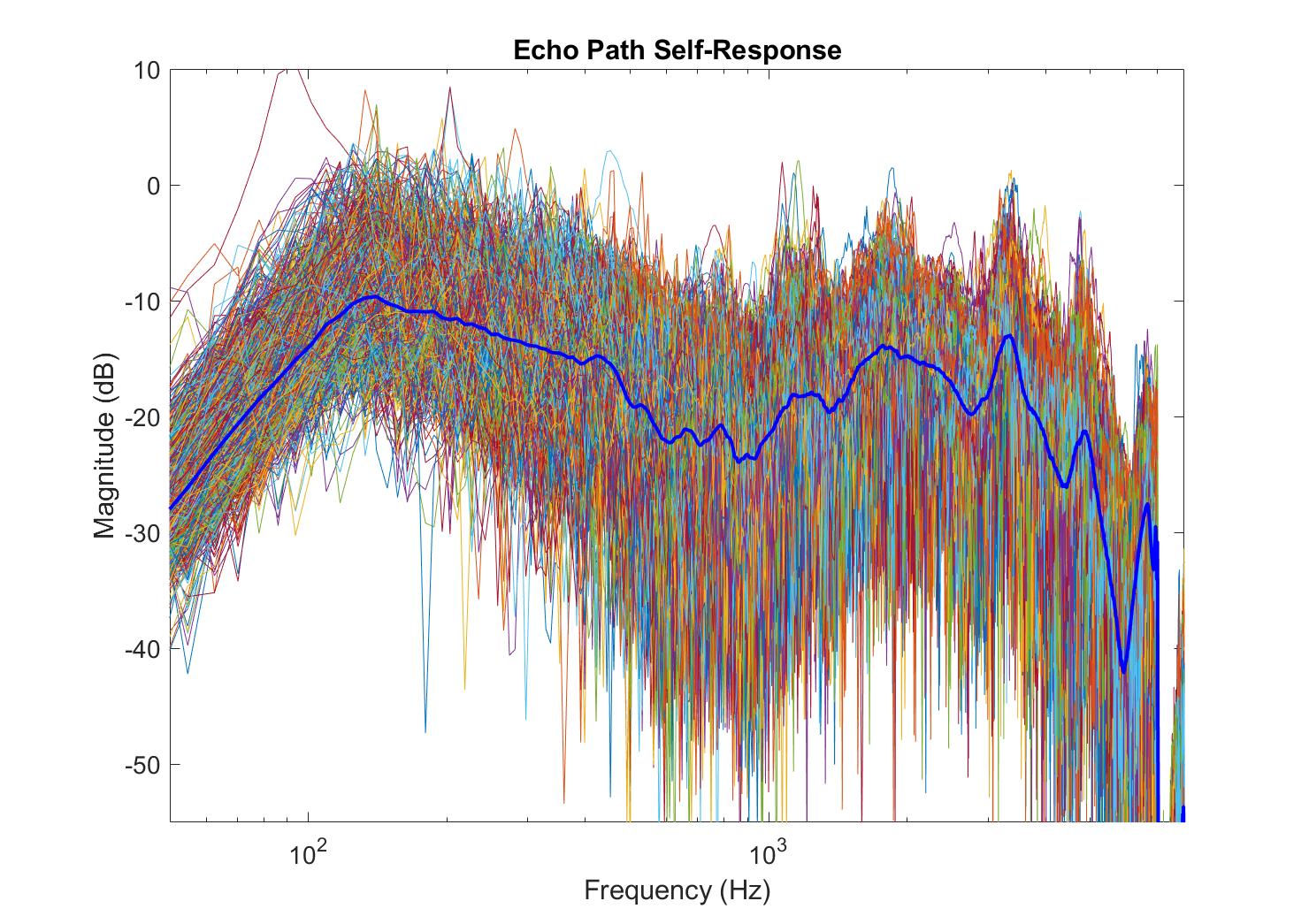}}\subfigure{
\hfill
\includegraphics[width=.5\linewidth]{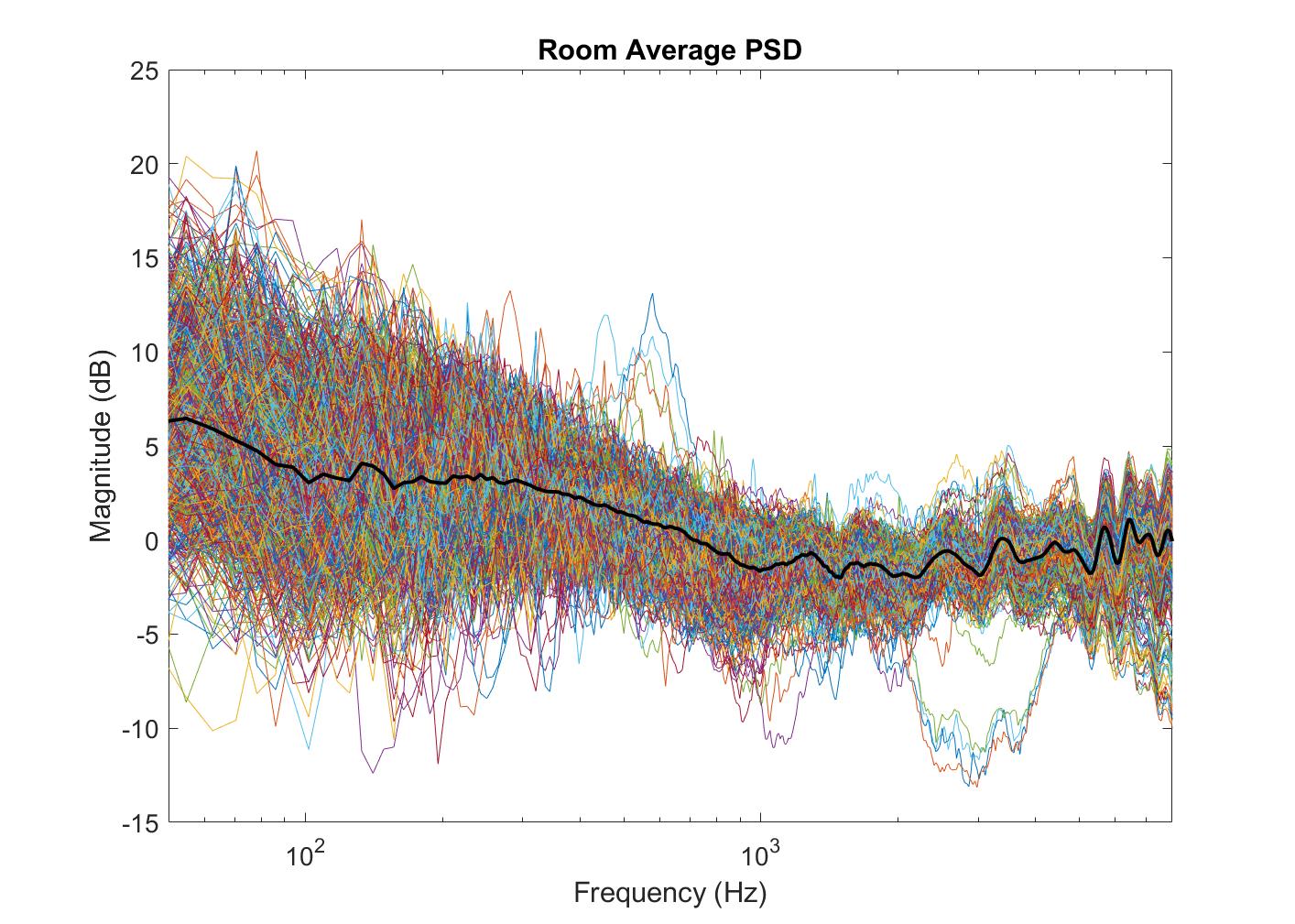}}
\vspace{-0.2cm}
\caption{Echo path self-responses and room average power responses from the collected database. Bold lines denote ensemble average.}
\vspace{-0.4cm}
\label{fig:FIG2}
\end{figure}
\begin{figure}[t]
\centering  
\subfigure{
\includegraphics[width=.5\linewidth]{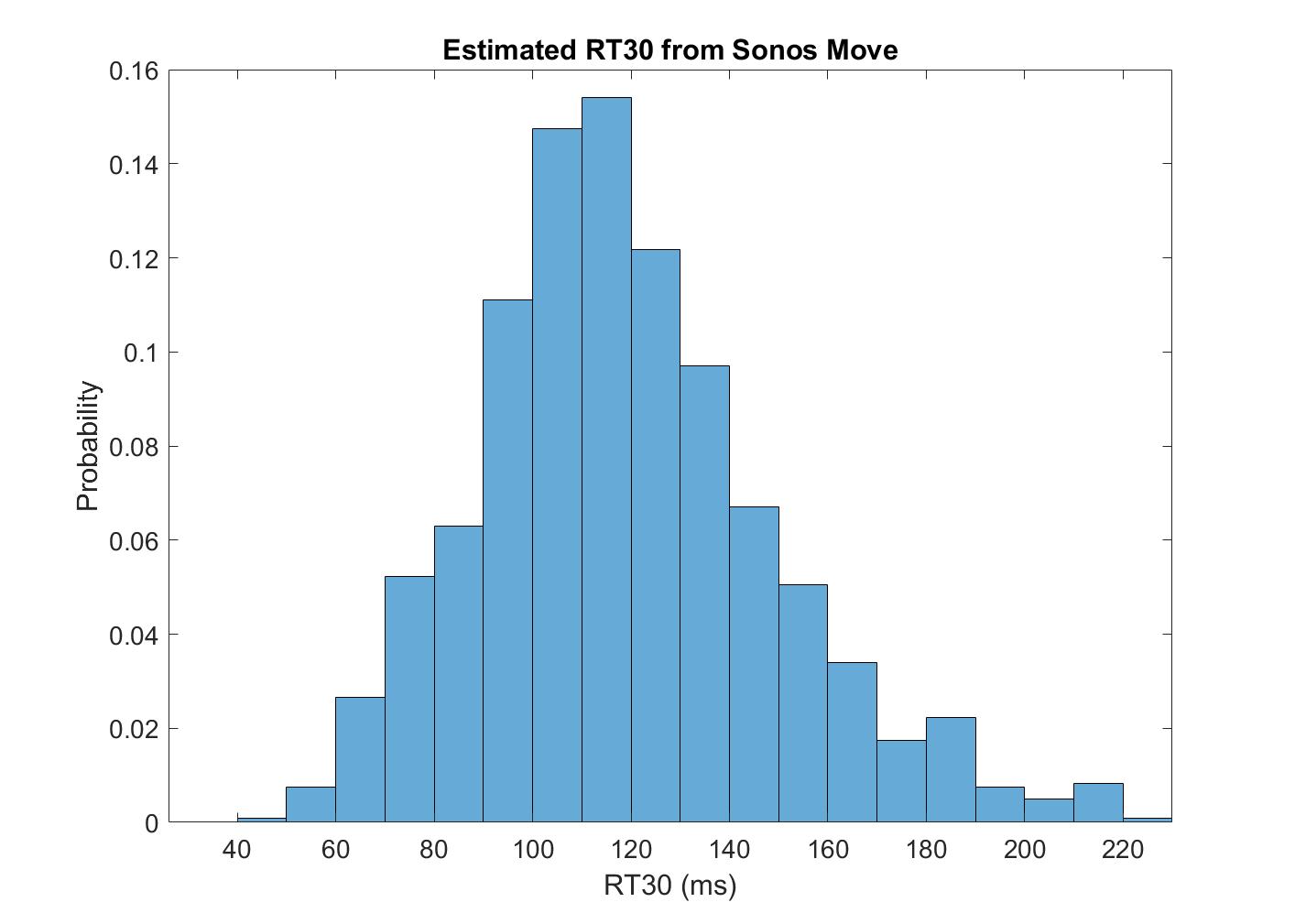}}\subfigure{
\hfill
\includegraphics[width=.5\linewidth]{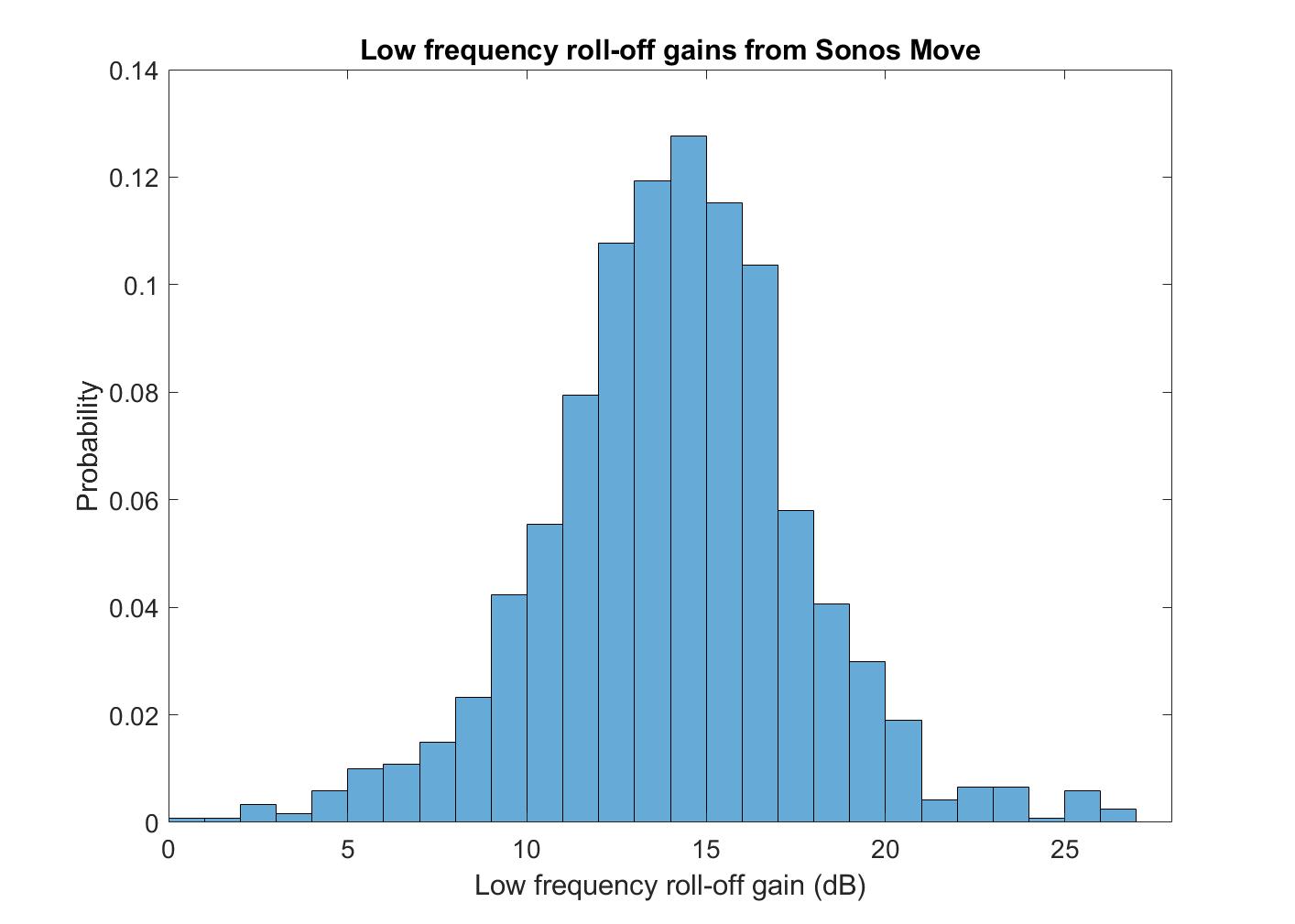}}
\vspace{-0.2cm}
\caption{The distribution of RT30 and low-frequency roll-off gains of echo path self-responses for Sonos Move database.}
\label{fig:FIG3}
\end{figure}
For data collection of echo path self-response/room average power response sets, a Sonos Move speaker (with built-in microphones) and one pre-calibrated external microphone were used. Room average power responses were measured as described in Sec.~\ref{sec:intro} and echo path self-responses were derived by playing sine sweep signals from the speaker. The Sonos Move speaker was placed in different rooms with a wide diversity of room types and at various locations in rooms (e.g. center, corner, near walls and etc.). Overall, as shown in Fig.~\ref{fig:FIG2}, the database includes a total of 1207 echo path self-response/room average power response sets at a sampling rate of 16 kHz and FFT length $N_{f}=2048$. The distribution histograms of the two investigated local features of echo path self-responses in Sec.~\ref{subsec:3:3}, i.e. RT30 and low-frequency roll-off gains, are demonstrated in Fig.~\ref{fig:FIG3}. From Fig.~\ref{fig:FIG3}, it can be seen that  both features follow a normal distribution well. Therefore, we fitted them with two normal distribution models and used the derived mean and standard deviation values to define the local subgroup division schemes for each of the features shown in Table 1.
\begin{table}[]
\centering
\vspace{-0.5cm}
\caption {Local subgroup division schemes} \label{tab:title} 
\vspace{0.2cm}
\begin{tabular}{|l|c|c|}
\hline
        & RT30                        & Low-freq roll-off           \\ \hline
Group 1 & {[}0,87) ms, 265 sets                  & {(}$-\infty$,10) dB,211 sets                    \\ \hline
Group 2 & {[}87,151{]} ms, 679 sets               & {[}10,18{]} dB, 758 sets              \\ \hline
Group 3 & (151,+$\infty$) ms, 263 sets  & (18,$\infty$) dB, 208 sets \\ \hline
\end{tabular}
\vspace{-0.5cm}
\end{table}
When evaluating the performance we used a repetitive cross-validation approach, i.e., we used data from $N_{tr}$ echo path self-response/Room average power response sets for training then evaluated the performance for $N_{val}$ unseen sets and repeated this process for 50 times. Note that the training data size for the local PCA-based method was $N_{tr}/3$ for each of the subgroup.
\vspace{-0.2cm}
\subsection{\label{subsec:4:2}Estimation of room average power response}
To evaluate the performance of the room average power response estimators, we used the following measure to quantify the dB level error between the estimated room average power response $\hat{r}(\omega)_{j}$ and the ground-truth ${r}(\omega)_{j}$ for the $j$th sample across frequencies:
\begin{equation}
\epsilon(\omega)_{j}=|10\log(|{r}(\omega)_{j}|^2)-10\log(|\hat{r}(\omega)_{j}|^2)|.
\end{equation}
\begin{figure}
\vspace{-0.4cm}
\centering
\includegraphics[width=0.65\linewidth]{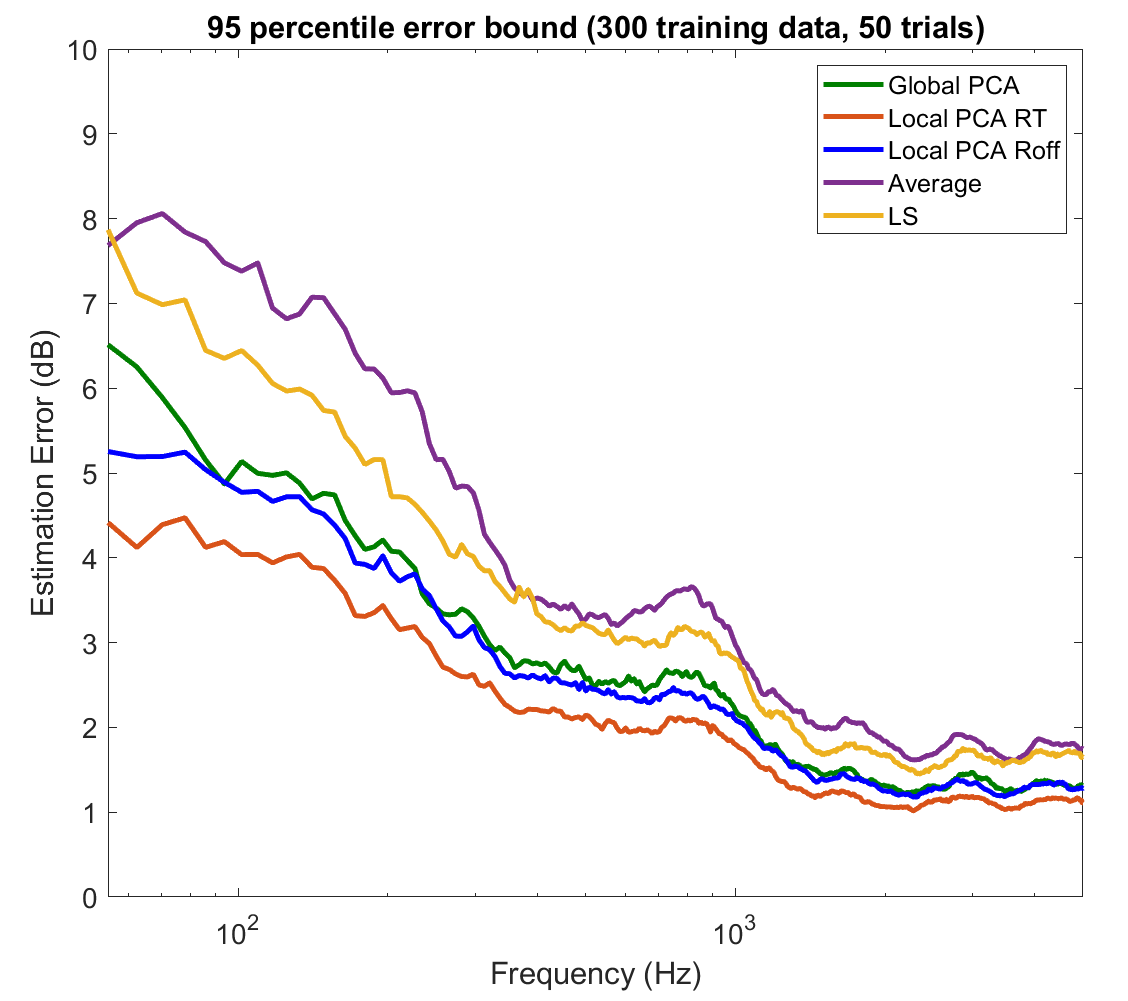}
\vspace{-0.3cm}
\caption{\label{fig:FIG4}95 percentile estimation error bounds of room average power response estimators using a repetitive cross-validation.}
\end{figure}
\begin{figure}[t]
\centerline{\includegraphics[width=0.65\linewidth]{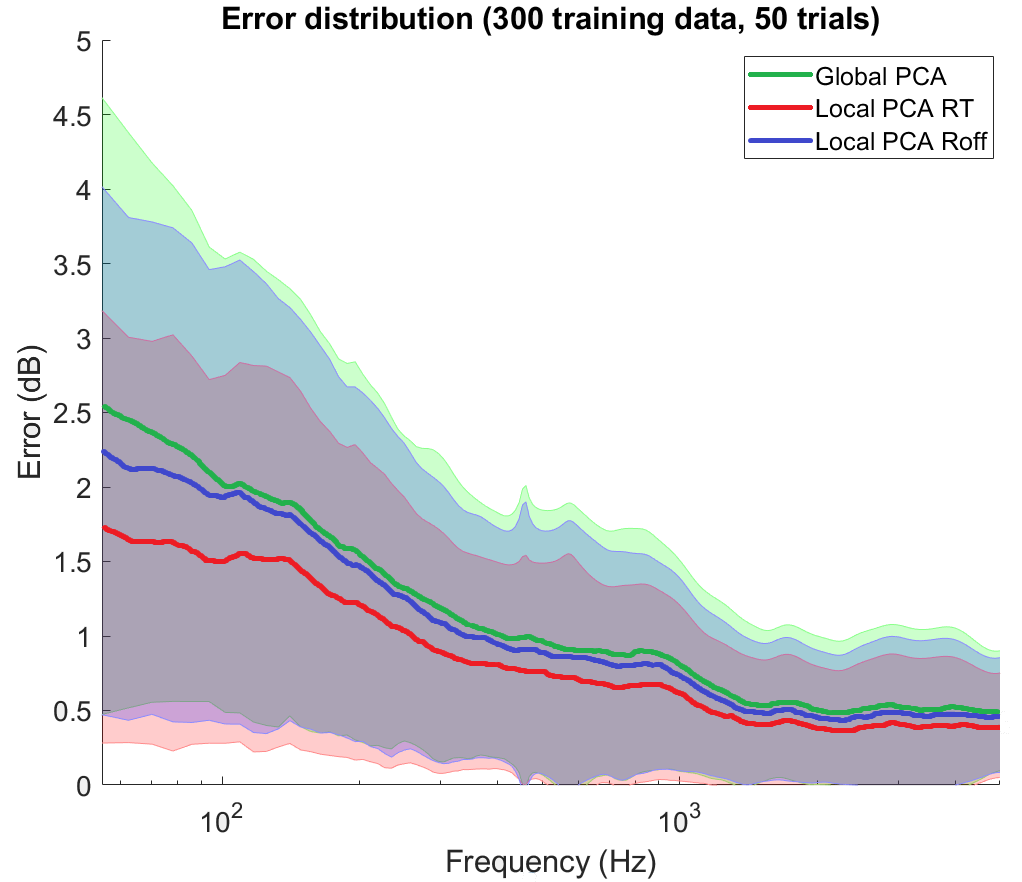}}
\vspace{-0.2cm}
\caption{\label{fig:FIG5}Average estimation error (bold lines) for three PCA-based estimators. Shaded areas indicate estimation error standard deviation.}
\vspace{-0.4cm}
\end{figure}
In particular, we focused on the 95th percentile error in dB level of the repetitive cross-validation process as we aimed to check if the majority of estimation processes deliver sufficient accuracy. In this evaluation, we chose $N_{tr}=300$ and $N_{val}=100$. In Fig.~\ref{fig:FIG4}, estimation performance of PCA-based systems (including both the ``Global PCA" method and local PCA methods ``Local PCA RT" and ``Local PCA Roff" that represents systems based on local features of RT30 and Low-frequency roll-off gain, respectively) is compared with the linear LS regression estimator (``LS") in Sec.~\ref{subsec:3:1} and the ensemble average estimator "Average". For all PCA-based estimation methods, we selected a fixed value of $K_r=32$ as we found that it consistently reconstructed raw room average power response coefficients with sufficient accuracy using only the first 32 orders of principal components. Note that the system of equations in Eq.~\eqref{eq:EQ2} is under-determined if $K_{s} < K_{r}$, hence, it is important to choose $K_{s} \ge K_{r}$. Therefore, we choose $K_{s}=240$ for the global PCA method and $K_{s}=80$ when training individual local PCA models.  

From Fig.~\ref{fig:FIG4}, it shows that a noticeably improved estimation performance can be achieved by all PCA–based approaches compared to the linear LS regression method using the same training dataset, which verifies advantages of the PCA due to its efficiency and dimensionality reduction. Both of the local PCA methods also outperform the global PCA method. Especially, "Local PCA RT" achieves an 95th percentile error bound that is within 5 dB across all frequencies. It clearly shows that the division of the global training dataset into sub-groups based on RT30 facilitates trained PCA models to capture the variability of acoustic impulse responses due to local reverberation conditions more accurately, which leads to superior estimation performance in comparison to the system using low-frequency roll-off gains as local features. Fig.~\ref{fig:FIG5} presents error regions across frequencies for all three PCA-based estimation schemes, which demonstrates the same performance pattern as in Fig.~\ref{fig:FIG4}. For the remaining part of the paper, the ``local PCA RT" system is referred to as the local PCA-based method.
\vspace{-0.2cm}
\subsection{\label{subsec:4:3} Acoustic room compensation}
To investigate how well the estimate of the room average power response $\hat{r}(\omega)$ leads to a design that matches the reference ``Trueplay" system with the ground-truth room average power response, we conducted a MUSHRA-style headphone listening test \cite{MUSHRA} with 11 listeners, including 6 professionally/self-reported trained listeners. For each listener, we randomly selected 4 self-response/room average power response sets from the validation dataset to simulate signals. Three stimulus signals were considered (including vocal music, jazz music and movie clips), which were then presented over a pre-equalized Beyerdynamic DT-880 headphone at 75 dBSPL.

In this evaluation, the following processing conditions are included: 1) Reference system with ground-truth room average power responses; 2) Stock system without room compensation processing; 3) ``LS300", denotes a filter design system using linear LS regression estimator with $N_{tr}=300$; 4)``LSMove", denotes the system using linear LS regression estimator with all available training data from the Sonos Move database, i.e. $N_{tr} \approx 1000$; 5)``GPCA", denotes the system using global PCA-based estimator with $N_{tr}=300$; 6)``LPCA300", denotes the system using local PCA-based estimator with $N_{tr}=300$; 7)``LPCA150", denotes the system using local PCA-based estimator with $N_{tr}=150$. Subjects were asked to rate the perceived sound similarity in comparison to provided reference samples.

\begin{figure}[t]
\centerline{\includegraphics[width=.7\linewidth]{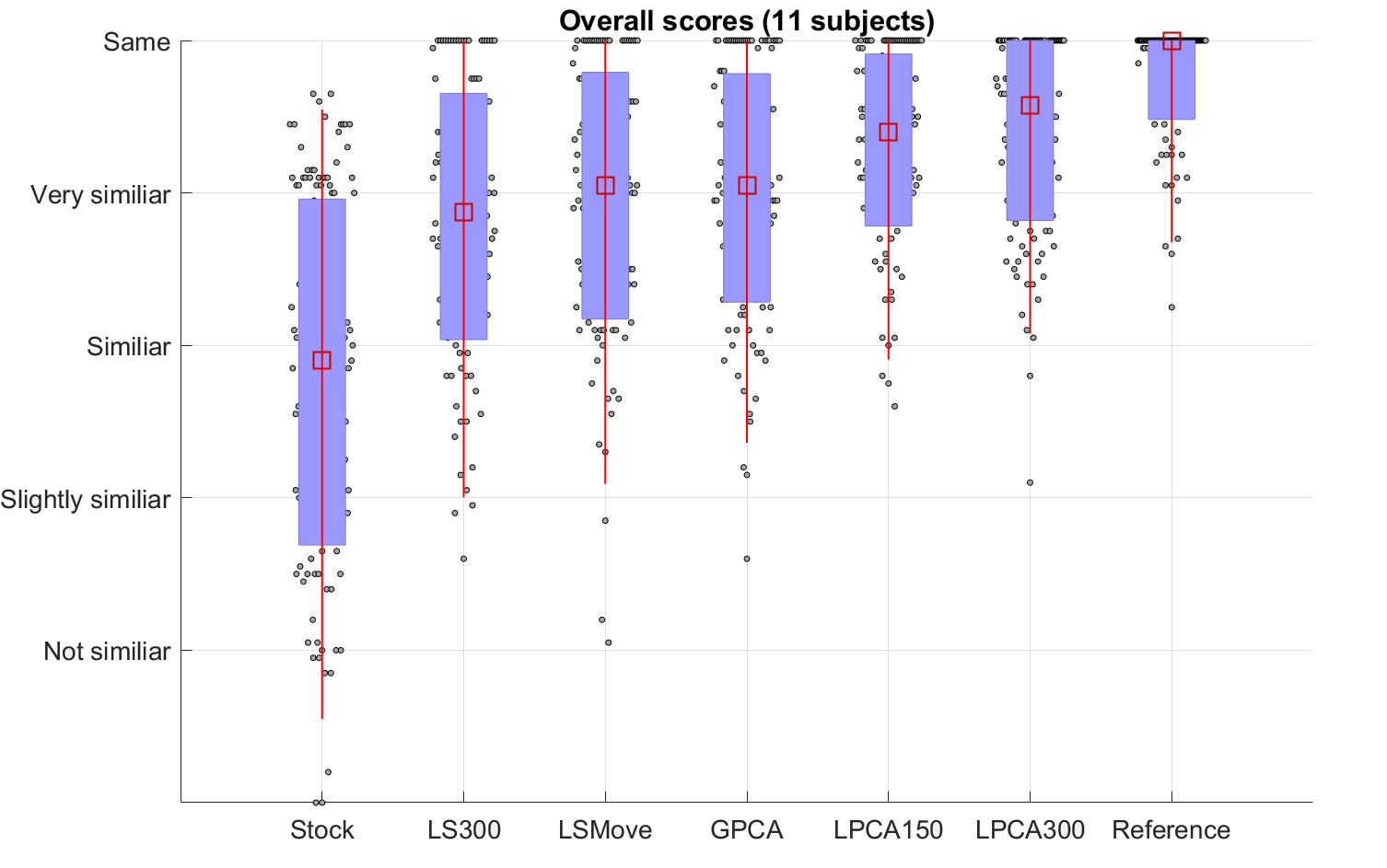}}
\vspace{-0.2cm}
\caption{\label{fig:FIG6}Sound similarity ratings of the MUSHRA study evaluating room compensation filters from all 11 listeners.}
\vspace{-0.3cm}
\end{figure}

\begin{figure}[t]
\centerline{\includegraphics[width=.7\linewidth]{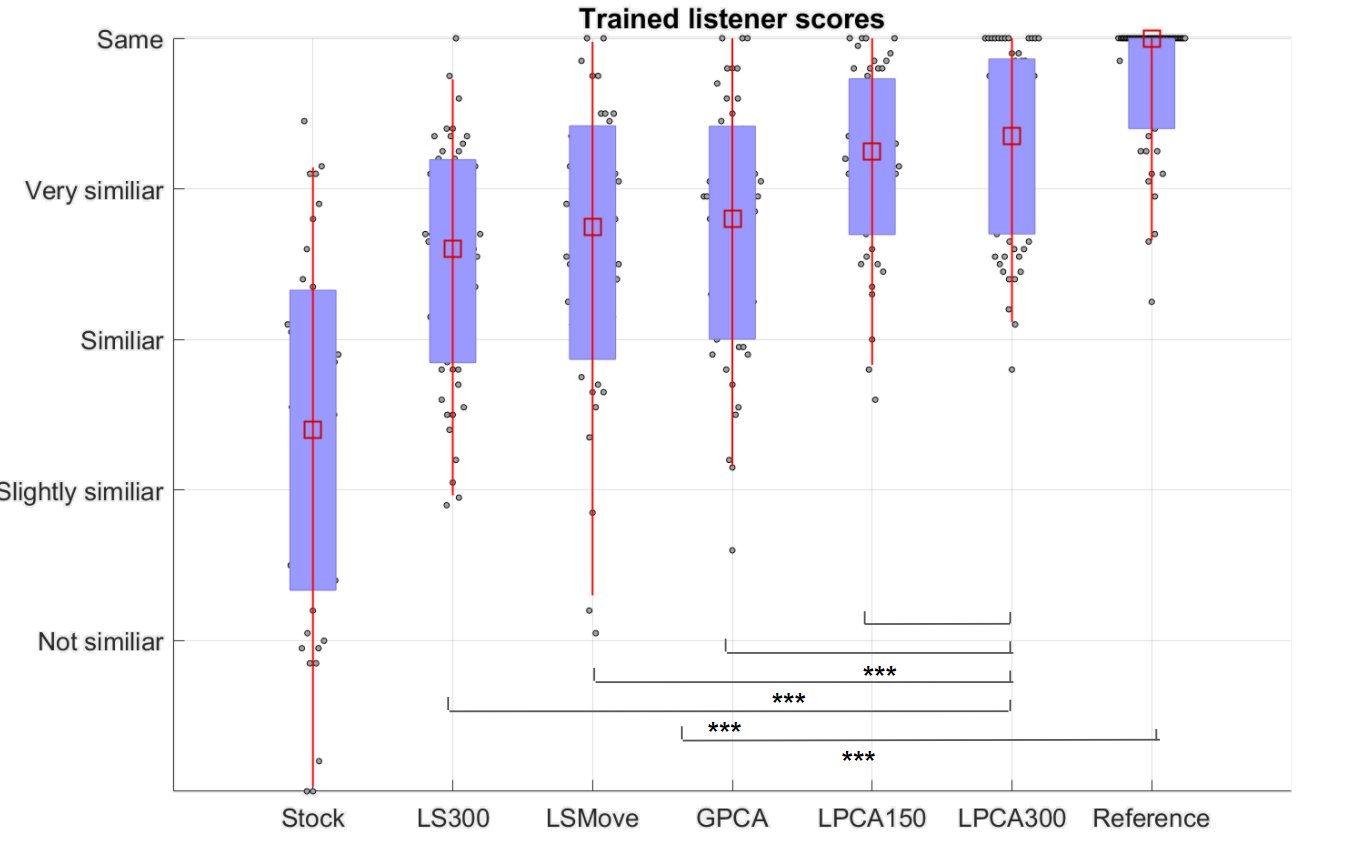}}
\vspace{-0.2cm}
\caption{\label{fig:FIG7}Sound similarity ratings of the MUSHRA study from 6 trained listeners (*** indicates p $<$ 0.001).}
\vspace{-0.5cm}
\end{figure}

Fig.~\ref{fig:FIG6} presents results of the MUSHRA test from all 11 subjects, which shows that most subjects were able to identify the hidden reference signals. However, several na\"ive listeners reported having difficulty distinguishing multiple samples that sounded highly similar. For trained listeners, results in Fig.~\ref{fig:FIG7} show that the stock condition was rated least similar to the reference. The remaining six conditions were analyzed using a linear mixed effects model \cite{lmmgui}. Apart from hidden references, highest similarity ratings were obtained by the proposed "LPCA300" system ($p<0.001$ compared to ``GPCA", ``LS300" and even ``LSMove" that uses a significantly larger training dataset). In addition, ``LPCA150" produces a close performance to "LPCA300" with $p>0.001$, which suggests that there is a potential to further reduce the size of the training dataset while maintaining comparable perceptual improvements for the PCA-based method. Overall, listening test results validates the proposed local PCA-based room average power response estimation approach.
\vspace{-0.3cm}
\section{Conclusion}
\label{sec:conclusion}
\vspace{-0.2cm}
In this work, a room compensation design by leveraging an estimate of room average power response is proposed. We present a PCA-based approach to predict room average power responses using measured echo path self-response at the on-board microphone. By exploiting reverberation time of echo path self-responses and deriving individual local PCA models that capture the local variability of impulse responses, the proposed method obtains an 95th percentile estimation error within $5$ dB, which facilitates the proposed room compensation filter to deliver a good sound similarity compared to the reference system. A MUSHRA-style listening test was conducted and results confirm the effectiveness of the approach.


\bibliographystyle{IEEEbib}
\bibliography{strings}

\end{document}